\documentclass[preprint,prb,showpacs]{revtex4}

\usepackage{graphicx}
\usepackage{dcolumn}
\usepackage{float}
\usepackage{amsmath}

\newcommand{\comment}[1]{}

\begin{document}

\title{\boldmath Response to comment by Cuk {\it et al.} \unboldmath}

\author{$^1$T. Valla, $^1$P.D. Johnson, $^1$Genda Gu, $^2$J. Hwang, and $^2$T. Timusk
} \email{timusk@mcmaster.ca}

\affiliation{$^{1}$Department of
Physics, Brookhaven National Laboratory, Upton, New York 11973-5000,
USA \\ $^{2}$Department of Physics and Astronomy, McMaster
University, Hamilton, ON L8S 4M1, Canada} \date{\today}

%\pacs{74.25.Gz, 74.62.Dh, 74.72.Hs}

%%%%%%%%%
%Abstract
\begin{abstract}In a recent study of the optical conductivity of the Bi-2212 high
temperature superconductor, Hwang, {\it et al.} have confirmed the
mass-renormalization changes observed in ARPES but find that such
changes are no longer observed in the highly overdoped regime. In a
comment on this study, Cuk, {\it et al.} challenge this conclusion and
present ARPES evidence for a kink in the superconducting state in the
highly overdoped regime. We show, however that two doping dependent
properties in the data of Cuk {\it et al.}, the frequencies of the
superconducting gap and the Van Hove singularity, suggest a $T_c$ that
is higher than the 58 K quoted, and that their sample is not as highly
overdoped as claimed. As a result, these data can not refute the claim
of Hwang {\it et al.} that the kink is absent in the highly overdoped
region.
\end{abstract}
%%%%%%%%%
\maketitle

%%%%%%%%%%%%%%%%%%%%%%
% introduction & body
%%%%%%%%%%%%%%%%%%%%%%

Since its initial observation \cite{valla99}, there has been
continued debate as to the origin and temperature dependence of a
mass-renormalization evident in the cuprates.  A change in the
magnitude of the renormalization on entering the superconducting
state, has emerged as a phenomenon that could help in identifying
any coupling mechanism \cite{kaminski01,lanzara01,johnson01}. In a
recent optical conductivity study, Hwang, {\it et al.}
\cite{hwang04} have confirmed the mass-renormalization changes
observed in ARPES \cite{johnson01}. Importantly, these authors
highlight the fact that such changes are no longer observed in the
highly overdoped regime. In a comment on the latter study, Cuk,
{\it et al.} \cite{cuk04} suggest that this conclusion is
unfounded and present ARPES evidence for the presence of a kink
developing in the superconducting state in the highly overdoped
regime. Here we examine the latter claim in detail.

Cuk, {\it et al.} \cite{cuk04} show ARPES data recorded in the
($\pi$, 0) region from an overdoped sample ($T_c$ = 58 K). Such
data shows two characteristic doping dependent properties well
documented in the literature, a gap in the superconducting state
that gets smaller with increased doping and the bonding state Van
Hove Singularity (VHS), corresponding to the binding energy at the
($\pi$,0) point. With increased doping, the VHS shifts to lower
binding energy.  In Fig. 1 we show the measured superconducting
gap (circles) at the Fermi surface in the vicinity of the
($\pi$,0) region.  The data are taken from our own studies
\cite{valla04} and the group of Shen \cite{feng01}. We also show
the gap measured in the cited study of Gromko {\it et al.}
\cite{kaminski03}.  The Gromko data would appear to have a larger
gap than one would anticipate for a sample with $T_c$ = 58 K. In
particular, the gap for their "58 K" sample is larger than the gap
reported by the Shen group \cite{feng01} for an overdoped sample
with $T_c$ = 65 K. The discrepancy becomes even more striking if
one examines the energy of the VHS.  In Fig. 1 we show (triangles)
a compilation of different measurements of the VHS
\cite{feng01,kaminski03,kim03}. The studies all show the same
reduction in binding energy with increasing doping.  The data of
Gromko {\it et al.} \cite{kaminski03} cited in the accompanying
comment is again inconsistent with this trend. Indeed, both the
measured superconducting gap and the VHS would indicate a $T_c$
higher than the quoted 58K, and probably higher than the first
neighboring point with $T_c$ of 65 K measured by Feng {\it et al.}
\cite{feng01}. A shift in doping level due to the loss of oxygen
is common for overdoped samples and such an effect will certainly
be present at the doping level of Gromko {\it et al.}'s
\cite{kaminski03} sample. We are left with the conclusion that the
comment of Cuk {\it et al.} \cite{cuk04} does not provide any
experimental evidence that refutes the claims in the paper of
Hwang {\it et al.} \cite{hwang04}.

Finally we make the observation that the change in the mass
renormalization that accompanies the superconducting transition
weakens beyond the detection limits in the overdoped regime for
doping levels in excess of approximately 0.23 - 0.24 in both
optical conductivity and photoemission. In fact a similar
conclusion has been made in the recent photoemission study of Kim
{\it et al.} \cite{kim03} from the ($\pi$,0) region.

% {\bf Acknowledgments}

\begin{figure}[t]
  \vspace*{+7.0cm}%
  \hspace*{+4.0cm}%
  \centerline{\includegraphics[width=3.5 in]{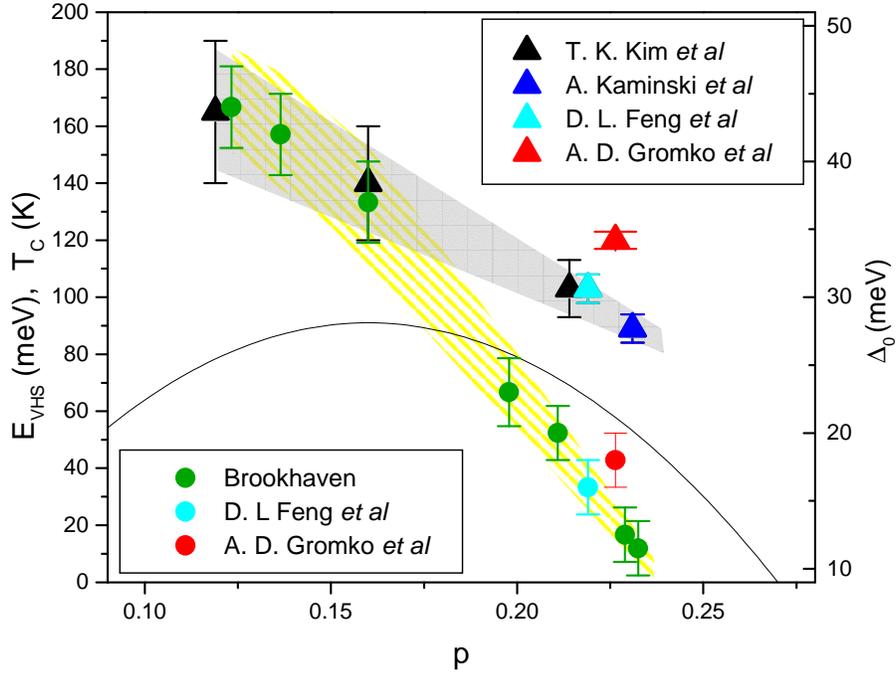}}%
  \vspace*{-10.5cm}%
  \caption{
The superconducting gap, $\Delta_0$, measured at ($\pi$,$k_{F}$)
(circles) and the energy of the bonding state, $E_{VHS}$, measured
at ($\pi$,0) (triangles) as a function of doping for Bi2212. Green
points have been measured at Brookhaven, cyan are from Feng {\it
et al.} \cite{feng01}, black triangles are from Kim {\it et al.}
\cite{kim03} and blue triangle is from Kaminski {\it et al.}.
\cite{kaminski01} Red data points are from Gromko
\cite{kaminski03}.}
  \label{Fig1}
\end{figure}

\end{document}